\documentclass[manuscript]{acmart}
\AtBeginDocument{%
  }

\usepackage{graphicx}
\usepackage{caption}
\usepackage{subcaption}
\usepackage{wrapfig}
\usepackage{xcolor}
\usepackage{booktabs, makecell, multirow, array}
\usepackage{wasysym}
\begin{document}

\title{Akita: A High Usability Simulation Framework for Computer Architecture}

\author{Sabila Al Jannat, Ying Li, Mengyang He, Xuzhong Wang, Huizhi Zhao, Jingxiang Sun, Daoxuan Xu, Enze Xu, Yifan Sun}
\email{{sajannat, yli81, mhe03, xwang58, hzhao10, jsun20, dxu05, exu03, ysun25}@wm.edu}
\affiliation{%
  \institution{William \& Mary}
  \city{Williamsburg}
  \state{VA}
  \country{USA}
}

\renewcommand{\shortauthors}{Jannat et al.}

\newcommand{\takeaway}[1]{
  \noindent\rule{\linewidth}{0.1pt}
  \par\nobreak\noindent\textbf{Takeaways:}
  #1
  \vspace{-2mm}
  \par\nobreak\noindent
  \rule{\linewidth}{0.1pt}
}

\newcommand{\yifan}[1]{\textcolor{red}{yifan: #1}}

\begin{abstract}
  Computer architecture simulation is essential for evaluating new designs without the need for costly tapeout. The community has developed dozens of valuable simulators that have enabled significant architectural advances. However, using and developing simulators remains a major barrier due to ad-hoc component interfaces, strict deployment requirements, the burden of managing performance optimizations like parallelization at the component level, and limited monitoring and visualization capabilities. The root cause of these limitations is the systematic neglect of user and developer experience in favor of technical functionality. We believe that only by separating technical concerns from user and developer experience concerns---through a dedicated simulation engine decoupled from hardware models---can the community overcome these fundamental obstacles and enable more productive architectural research.

  Akita embodies this philosophy as a dedicated simulation engine that cleanly separates infrastructure from architectural models. Smart Ticking and Availability Backpropagation let developers write simple cycle-based code while achieving event-driven performance. Parallel simulation happens transparently---developers write single-threaded code while Akita handles multi-core execution. Akita's simple, uniform, yet powerful simulation tracing support enables real-time monitoring and post-simulation visualization. We demonstrate the flexibility of Akita through case studies, including the development of a trace-based DNN simulation and a RISC-V CPU simulation, showing how prioritizing developer experience accelerates architectural research.
\end{abstract}






\maketitle

\section{Introduction} \label{sec:intro}

Computer architecture simulators are essential tools that facilitate the development of new chips and computer systems~\cite{akram2019survey}. Compared to on-silicon evaluation, simulation enables researchers to study a wide range of design configurations and ideas without requiring costly hardware prototypes, significantly accelerating the design cycle.

As the demand for innovative chip and computer system designs escalates, a diverse array of computer architecture simulators has emerged. These simulators cover a broad spectrum, targeting CPUs~\cite{wenisch2006simflex,binkert2011gem5,patel2011marss,carlson2011sniper}, GPUs~\cite{bakhoda2009analyzing,ubal2012multi2sim,power2014gem5,sun2019mgpusim,khairy2020accel,rashidi2020astra,li2025triosim}, memory systems~\cite{uhlig1997trace,wang2005dramsim,kim2015ramulator}, interconnects~\cite{jiang2010booksim,abad2012topaz}, and domain-specific accelerators~\cite{cong2014architecture,munoz2021stonne}. While most are specialized for particular domains, a few (e.g., gem5~\cite{binkert2011gem5}, SST~\cite{rodrigues2011structural}, and Multi2Sim~\cite{ubal2012multi2sim}) stand out for their ability to function as simulation engines (software frameworks that provide common requirements for building simulators around hardware models).

% \begin{figure}[t!]
%  \centering
%  \includegraphics[width=0.7\linewidth]{images/engine.drawio.pdf}
% \caption{The proposed engine-centric simulator development model breaks down monolithic simulator repositories, bringing benefits including enhanced user/developer-friendliness, reduced repeated optimizations, and easier contribution to the community.}
% \label{fig:engine}
% \end{figure}

\begin{wrapfigure}{r}{0.4\linewidth}
  \centering
  \includegraphics[width=\linewidth]{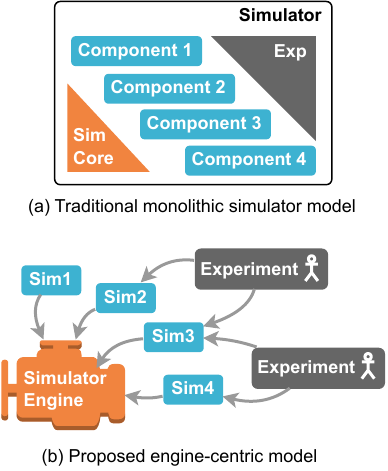}
  \caption{The proposed engine-centric simulator development model breaks down monolithic simulator repositories, bringing benefits including improved user and developer friendliness, reduced repeated optimizations, and easier community contributions.}
  \label{fig:engine}
\end{wrapfigure}

Simulation engines provide core functionality such as time advancement (event-driven or cycle-based), component state management, and inter-component communication. Some also offer usability-enhancing features such as execution trace visualization~\cite{ariel2010visualizing,ziabari2015visualization} and configurable debugging tools. Because simulators rely heavily on the underlying engine, the engine's design directly shapes developers' efficiency and users' ability to analyze results.
% However, most existing simulation engines have been designed primarily with a focus on technical functionality, often overlooking usability, extensibility, and developer-friendliness.

Many widely used simulators adopt a monolithic design, where experiment configurations and architectural models are tightly coupled with the simulation core. The monolithic design merges technical concerns (e.g., covering every component, estimating execution time with low average error) with usability concerns (e.g., how developers can add a cache model without understanding the entire repository). Consequently, the primary metric becomes ``it works,'' regardless of how painful it is to use or extend.
This technical-first approach creates severe barriers: developers navigate 500K+ lines to add simple components, users install massive frameworks for basic experiments, and neither can share nor combine components across simulators.
Meanwhile, monitoring and visualization remain afterthoughts, leaving researchers blind during multi-day simulations and slowing architectural insights.
We argue that these challenges must be addressed at the engine level. By cleanly separating the simulation engine from architectural models and user experiments, we can improve modularity, enhance usability, and build a more sustainable foundation for community-driven development (see Figure~\ref{fig:engine}).

We introduce the Akita simulation engine\footnote{Akita is open-sourced at https://github.com/sarchlab/akita.}, which directly addresses these barriers through a radically different approach. Akita extracts the common requirements of all simulators, including time advancement, component communication, metrics collection, tracing, monitoring, and visualization, into a standalone, architecture-agnostic engine. This separation enables us to optimize infrastructure once while developers focus on modeling: they write components against simple APIs, users compose simulators from interchangeable parts, and everyone benefits from transparent parallelization and built-in monitoring. By treating simulation infrastructure as a first-class concern separate from architectural models, Akita reduces development friction and provides comprehensive tools for understanding hardware behavior, helping researchers move from implementation struggles to architectural insights.

Akita aims to achieve the technical goals of a computer architecture simulation engine while balancing developer and user experience. We highlight a set of features that enhance simulators with minimal or no additional developer burden. Key features of Akita include:
\begin{itemize}
  \item \textbf{Smart Ticking with Availability Backpropagation} combines the simplicity of cycle-based simulation with event-driven efficiency by automatically skipping unnecessary component ticks when no progress can be made, achieving an average 2.68$\times$ speedup. Availability Backpropagation is the underlying mechanism that enables Smart Ticking by propagating buffer availability information backward through ports and connections, waking up components only when they can make forward progress.
  \item \textbf{Transparent parallel simulation} enables parallel execution of simulations using multiple CPU cores (achieving 1.88$\times$--2.38$\times$ speedup) without requiring developers to modify their single-threaded component implementations.
  \item \textbf{Task-based tracing system} provides an aspect-oriented tracing infrastructure that tracks hierarchical tasks with parent-child relationships, requiring minimal code instrumentation while collecting comprehensive performance metrics and enabling enhanced debugging backtraces.
  \item \textbf{Real-time monitoring with AkitaRTM} offers live dashboards and analysis tools to monitor simulation progress, identify bottlenecks, and debug hangs without restarting simulations.
  \item \textbf{Post-simulation trace visualization with Daisen} automatically generates detailed execution traces for examining cycle-by-cycle hardware behavior after simulation completion.
\end{itemize}

% Overall, Akita aims to be an architecture-agnostic simulation engine that focuses on improving developer and user friendliness at the engine level, benefiting all simulators built on top of it. This paper makes the following contributions:

% \begin{itemize}[leftmargin=*, nosep]
%     \item We present Akita, a simulation engine that streamlines the development of individual computer architecture simulators, emphasizing user and developer friendliness.
%     \item We introduce Smart Ticking and Availability Backpropagation as methodologies that allow cycle-based modeling with the efficiency of event-driven simulation, combining simplicity with high performance.
%     \item We introduce a feature-rich tracing system that enables efficient metrics collection, real-time visualization, and improved simulator debugging and analysis.
%     % \item We demonstrate Akita's efficiency and usability through case studies and a developer study.
% \end{itemize}

\section{Motivation}\label{sec:goals}

Akita employs a holistic approach to analyzing simulator requirements across two dimensions: the simulator life cycle (configuration, deployment, execution, debugging, and result analysis) and stakeholder roles. Developers (marked \underline{DX}) create and integrate new hardware models, while users (marked \underline{UX}) configure systems, run experiments, and analyze results. Cross-cutting these five stages with the two roles yields 10 design requirements that shape Akita's development, ensuring the engine addresses not just technical functionality but the full spectrum of user and developer experiences throughout a simulator's life cycle.

\subsection{Configuration}

Simulator users tweak the system to explore their unique design space. Supporting flexible configurations is essential for a simulator.

\underline{UX-1}: \textbf{Structural flexibility.} Users should be able to reconfigure the organization of the system (e.g., placing address translation before, between, or after L1/L2 caches) by choosing and reconnecting components in configuration, rather than by changing engine or model code.

\underline{DX-1a}: \textbf{Protocol-first, extensible component design.} Developers implement components against small, stable protocol interfaces (for example, memory request/response messages) rather than against specific peer components. Following this principle reduces developer effort (by focusing on the protocol) and keeps components interchangeable and reusable.

\underline{DX-1b}: \textbf{Open--closed component design.} Following the open--closed principle~\cite{martin2009clean}, common policies (e.g., replacement, arbitration, address mapping) should be supplied via pluggable modules or configuration, so that adding a new option usually means adding a new policy module rather than editing the core component code (e.g., adding an \texttt{else} branch).

Modern simulators like gem5~\cite{binkert2011gem5,lowe2020gem5}, SST~\cite{rodrigues2011structural}, and Multi2Sim~\cite{ubal2012multi2sim} already provide strong support for modular components and flexible configuration: gem5 through Python/SimObjects hierarchies, SST through plug-in components with StandardMem protocols, and Multi2Sim through structured configuration files. These approaches work well for their respective communities.

Akita explores a complementary design point emphasizing: (i) protocol-first interfaces enforced at the engine level, ensuring all components implementing a protocol are truly interchangeable, and (ii) pervasive use of builder patterns with dependency injection, systematically separating component behavior from policies (replacement algorithms, arbitration strategies, routing decisions) and configuration parameters. While we provide first-party components for convenience, the same builder interface works uniformly for both internal and external components. These software engineering choices represent incremental refinements rather than fundamental breakthroughs, yet our evaluation shows they reduce configuration complexity while maintaining research flexibility. We therefore view Akita as exploring a different point in the modularity-simplicity design space, complementing rather than replacing existing engines.

\subsection{Deployment}

Simulators must run reliably across diverse development and compute environments, yet setup complexity often becomes a barrier to adoption.

\underline{UX-2}: \textbf{Low-friction setup.} Users should be able to install and run simulators with minimal system dependencies, avoiding version-sensitive toolchains or privileged installations, with binaries that work across machines without reconfiguration.

\underline{DX-2}: \textbf{Cross-platform development.} Developers should write once and deploy anywhere, using consistent tooling across different operating systems (Windows/macOS/Linux) and CPU architectures (x64/ARM/RISC-V) without platform-specific modifications.

Existing frameworks handle deployment with varying degrees of complexity. gem5~\cite{binkert2011gem5,lowe2020gem5} requires matching C++ compilers, Python environments, and system libraries across deployment targets; while containerization helps, users must still manage Docker or Singularity setups. SST~\cite{rodrigues2011structural} assumes an MPI stack appropriate for HPC clusters but heavyweight for laptop development. Multi2Sim~\cite{ubal2012multi2sim} primarily targets Linux environments with limited cross-platform support.

Akita leverages Go's toolchain for a fundamentally different deployment experience: static binaries with no runtime dependencies, built-in cross-compilation (e.g., \texttt{GOOS=linux GOARCH=arm64 go build}), and identical behavior across platforms. A researcher can develop on an ARM-based Mac, compile for powerful x64 Linux clusters, and share the resulting binary with colleagues who can run it immediately---no dependency installation, no version conflicts, no ``works on my machine'' issues.

\subsection{Execution}
The simulation performance gap continues to widen as hardware complexity grows~\cite{akram2019survey}, motivating techniques from sampling~\cite{carlson2011sniper,liu2023photon} to statistical modeling~\cite{li2023path,lee2025forecasting}. Beyond these algorithmic advances, simulation engines may provide structural optimizations (e.g., eliminating idle component polling, enabling parallel execution, and mixing abstraction levels) without burdening component developers.

\underline{UX-3}: \textbf{Mixed-mode simulation.} Users can combine multiple abstraction levels (cycle-accurate, functional, trace-driven, high-level) within one simulation, trading accuracy for speed where appropriate.

\underline{DX-3}: \textbf{Transparent optimization.} Developers write simple, single-threaded, cycle-based components while the engine automatically handles performance optimizations (idle skipping, parallel execution) without explicit threading or event management in component code.

Existing frameworks offer various optimization strategies. gem5~\cite{binkert2011gem5,lowe2020gem5} provides atomic/timing/detailed CPU models and dist-gem5~\cite{mohammad2017dist} for distributed simulation, though parallelization requires specific configuration and checkpointing rather than being transparent to component code. SST~\cite{rodrigues2011structural} delivers built-in parallel discrete-event simulation over MPI, but components must be written with PDES semantics in mind, which requires understanding event ordering, managing state consistency, and handling synchronization. Multi2Sim~\cite{ubal2012multi2sim} remains essentially single-threaded without general mechanisms for idle skipping or transparent parallelization.

These approaches work but place optimization burden on component developers. In contrast, Akita detects and skips idle components based on message availability, and the engine handles sleep--wake scheduling. Similarly, Akita's parallel execution requires no component modifications; the same single-threaded component code runs in parallel based on conservative timestamp synchronization at the engine level. While these techniques have similar precedents (e.g., HAsim's credits~\cite{pellauer2011hasim}, null message algorithms~\cite{davis1990distributed}), Akita packages them as transparent engine services, allowing developers to write straightforward cycle-based logic while achieving event-driven performance.

\subsection{Debugging}

Multi-day simulations that hang or crash after hours of execution represent a significant productivity drain, yet most debugging support remains offline (analyzing logs after failure) rather than online (monitoring during execution).

\underline{UX-4}: \textbf{Live monitoring and hang diagnosis.} Users can observe simulation health in real-time and identify stalled components or congested buffers without restarting multi-day runs.

\underline{DX-4}: \textbf{Architecture-aware debugging context.} Crashes and errors show the chain of architectural events (which instruction triggered which cache miss caused which memory transaction) rather than just software call stacks.

Simulator debugging approaches vary in sophistication. gem5~\cite{binkert2011gem5,lowe2020gem5} offers comprehensive \texttt{DPRINTF} tracing with debug flags and detailed statistics. Still, a hanging simulation typically requires restart with additional instrumentation. SST~\cite{rodrigues2011structural} and Multi2Sim~\cite{ubal2012multi2sim} similarly rely on logging and offline analysis.

Akita provides built-in runtime monitoring to detect hangs and bottlenecks in running simulations without restarting. When errors occur, task-based tracing shows the architectural cause chain (e.g., which instruction led to which cache miss led to which fault) rather than just software stack traces. Users report these capabilities significantly reduce debugging time for long-running simulations, particularly for intermittent bugs that would otherwise require multiple restart attempts to diagnose.

\subsection{Result Analysis}

Beyond aggregate performance metrics, understanding why architectural changes affect performance requires detailed execution traces and the ability to zoom between high-level trends and cycle-by-cycle behavior.

\underline{UX-5}: \textbf{Unified metrics and visualization.} Users can access both high-level statistics and detailed temporal traces from a single run, with built-in visualization tools for exploring performance without external scripting.

\underline{DX-5}: \textbf{Instrumentation separation.} Developers add metrics and traces through a uniform API that keeps measurement code separate from component logic, avoiding the common anti-pattern of performance counters scattered throughout hardware models.

Existing frameworks provide statistics APIs and performance counters—gem5~\cite{binkert2011gem5,lowe2020gem5} with hierarchical counters, SST~\cite{rodrigues2011structural} with multi-format output, Multi2Sim~\cite{ubal2012multi2sim} with domain-specific visualization, but share common limitations: metrics require direct component modification, visualization depends on external post-processing of text/CSV outputs, and instrumentation code remains intertwined with hardware logic.

Akita treats performance analysis as an engine-level concern with a uniform tracing API that works across all component types. Instrumentation points remain minimal in the component code while the engine handles collection, storage, and visualization based on configuration. This separation enables both real-time monitoring during simulation and post-simulation analysis without the traditional workflow of parsing large text logs offline.

\section{Key Features of Akita}

Next, we introduce Akita, focusing on how Akita features satisfy the UX and DX requirements for a simulation engine enumerated earlier. Akita uses the Go programming language to deliver a better developer experience without being significantly slower than C/C++, reflecting our prioritization of enhancing user and developer experience (satisfying \underline{UX-2} and \underline{DX-2}). A detailed comparison is listed in \autoref{tab:language_comparison}. With the growing popularity of using generative AI in programming, language is no longer a major porting barrier.

In addition to the engine-level features to be introduced, Akita ships with a wide range of first-party components, including caches with different write policies, DRAM modules ported from DRAMSim3~\cite{li2020dramsim3}, TLBs and MMUs, on-chip and off-chip network models (e.g., PCIe).

\newcommand{\good}[0]{\textcolor{green!60!black}{$\checkmark$}}
\newcommand{\ok}[0]{\textcolor{yellow!60!black}{$\circ$}}
\newcommand{\bad}[0]{\textcolor{red!60!black}{$\times$}}

\begin{wraptable}{r}{0.6\linewidth}
  \centering
  \vspace{-1cm}
  \caption{Comparison of Go and C/C++ for Simulator Development}
  \label{tab:language_comparison}
  \begin{tabular}{llll}
    \toprule
    \textbf{Feature} & \textbf{Lang} & & \textbf{Comment} \\
    \midrule
    Performance & Go & \ok & Sufficient for simulation \\
    & C/C++ & \good & Fast \\
    Simplicity & Go & \ok & Simple, but can be nuanced \\
    & C/C++ & \bad & Complex \\
    Dependency & Go & \good & One command package importing\\
    & C/C++ & \bad & Needs installation \\
    Cross-platform & Go & \good & Cross-compilation, static linking \\
    & C/C++ & \ok & Dynamic linking causes challenge \\
    Concurrency & Go & \good & Goroutines \& channels \\
    & C/C++ & \ok & pThreads/libs + boilerplate \\
    Memory & Go & \ok & Garbage collected, extra overhead \\
    & C/C++ & \ok & Manual, fast \\
    Comp. Speed & Go & \good & Fast builds, no config needed \\
    & C/C++ & \bad & Slower builds, CMake-like config \\
    Ecosystem & Go & \ok & New to arch. community \\
    & C/C++ & \good & Established in arch. community \\
    \bottomrule
  \end{tabular}
\end{wraptable}

\subsection{Organizations}

% \begin{figure}[t!]
%  \centering
%  \includegraphics[width=0.5\linewidth]{images/organization.drawio.pdf}
% \caption{Akita models the simulated hardware components using software components that collaborate by exchanging messages through ports and connections.}
% \label{fig:organization}
% \vspace{-0.5cm}
% \end{figure}

\begin{wrapfigure}{r}{0.5\textwidth}
  \centering
  \includegraphics[width=0.5\textwidth]{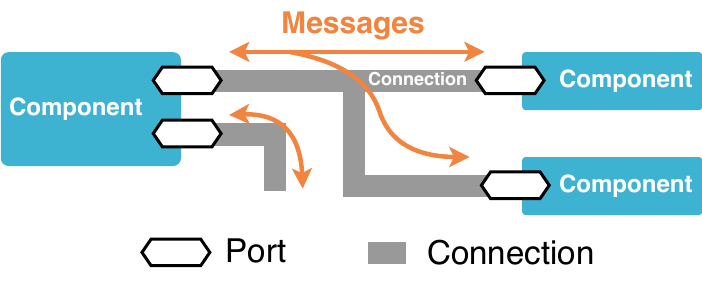}
  \caption{Akita models the simulated hardware components using software components that collaborate by exchanging messages through ports and connections.}
  \label{fig:organization}
  % \vspace{-0.5cm}
\end{wrapfigure}

Akita follows a similar system organization as used in many existing architectural simulators~\cite{binkert2011gem5, rodrigues2011structural}. This can help to reduce the learning curve for new users and developers. As shown in~\autoref{fig:organization}, key concepts include \emph{components}, \emph{ports}, \emph{connections}, and \emph{messages}.

\textbf{Component.} Simulator developers typically divide large systems into smaller, relatively independent elements (called \emph{components} in Akita), such as cores, caches, and memory controllers. Defining components is the first step towards a modular design, which serves to increase flexibility (satisfying \underline{UX-1}).

\textbf{Message.} Components communicate by exchanging messages. Messages are pure-data structures that include metadata (e.g., source and destination) and payload. Developers can define customized payload types that carry data.

% Akita defines two special types of messages as requests and responses. For example, a computing core sends a memory read request to a memory unit, and the memory unit can send data-ready responses back to the core. This specialization enhances simulation traceability (see~\autoref{fig:organization}).

Using messages formalizes the communication protocol between components. We do not allow function calls from one component to invoke another component, thus avoiding inconsistent per-component programming interface design, enhancing modularity, and better supporting requirement \underline{UX-1} and \underline{DX-1a}.

\textbf{Port.}
Components exchange messages through ports, each managing an incoming and outgoing buffer. Upon sending a message, the component calls the \texttt{Send} function of the port. If the port runs out of outgoing buffer space, it rejects the message, and the component can try to send the message later. Note that this mechanism often determines if a component can make forward progress (or not), which can be leveraged to enhance simulation performance (see~\ref{sec:smart_ticking} for more details).

Components can equip multiple ports for different communication purposes. For instance, a computing core may utilize one port to connect to the instruction cache and another to the data cache. Unlike gem5~\cite{binkert2011gem5}, Akita does not differentiate between master and slave ports, simplifying configuration (satisfying \underline{UX-1}).

\textbf{Connection.} Connections deliver messages from the source port to the destination port. A connection can be as simple as wires (delays are neglected) or as complex as full on-chip or off-chip network systems.

Unlike other simulators~\cite{binkert2011gem5, ubal2012multi2sim}, Akita allows a single connection to link multiple ports, although each port can only be served by one connection. When connecting multiple ports, the connection functions as a round-robin arbitrated crossbar, eliminating the need to design separate switches and simplifying configuration. However, if users require a specialized switch or arbitrator, Akita remains flexible; developers can integrate such components as needed (satisfying \underline{UX-1} and \underline{DX-1b}).

\subsection{Smart Ticking}~\label{sec:smart_ticking}

\textbf{Event-driven simulation.} Simulators typically use either cycle-based or event-driven strategies~\cite{akram2019survey}. Event-driven simulation prevents simulators from polling event components every cycle, allowing a component to fast-forward to more meaningful time ranges (i.e., when the component can make substantial progress). Additionally, event-driven simulation can be more flexible by scheduling events that are not on the cycle boundaries. These advantages drive the decision to employ a pure event-driven simulation mechanism in Akita, partially fulfilling \underline{UX-3}.

Despite the advantages of event-driven simulation, simply applying a pure event-driven simulation in computer architecture simulators has two problems.

1) Developers often find it difficult to implement event-driven simulators, as they need to define many event types and capture complex chronological ordering.

2) Pure event-driven simulation also experiences performance issues. For example, suppose we have a component that needs to perform an action that involves sending a message, but the message fails to be sent due to a full buffer. Since the component cannot predict when the buffer will free up, it can only schedule a retry event in every subsequent cycle until a buffer slot becomes available, almost falling back to cycle-based simulation.

\textbf{Cycle-based component interface.} We use Smart Ticking to address these two problems. Noticing that developers prefer to code in cycle-based style, we define a \texttt{Tick} event, which can be scheduled periodically by each component. Each component can define a \texttt{Tick} function to handle the tick event. By building the cycle-based interface above the event-driven foundation, we achieve higher flexibility, allowing developers to combine simulation schemes in a simulator or even within one component, fulfilling \underline{UX-3}.

\textbf{Rational behind Smart Ticking.} A critical challenge is preventing this event-based ticking method from falling back to cycle-based performance. We explain how we address this problem using a list of questions and answers.

\begin{itemize}
\item {\underline{Q}}: How can we improve the performance of ticking components based on Akita's event-driven simulation method? {\underline{A}}: We need to skip unnecessary ticks.
\item {\underline{Q}}: What ticks are safe to be skipped without impacting simulation accuracy? \underline{A}: If the tick is not making substantial progress, it is safe to skip it.
\item {\underline{Q}}: What causes a component to be unable to make progress? {\underline{A}}: Two situations - 1) when the component is fully idle (this includes the case if a component is waiting for a critical response, e.g., a core waiting for data); 2) when the component needs to send messages out, but the outgoing buffers are full.
\item {\underline{Q}}: What happens if a component does not make progress in a tick? {\underline{A}}:  If a component fails to make forward progress in a cycle, merely retrying will not change the outcome in future cycles. Progress will only resume if the underlying issue is addressed, such as the component becoming active (not idle) or having available space in the outgoing buffers. Thus, if a component is stalled, continuing the ticking process is unnecessary.
\item {\underline{Q}}: In what situation do we need to wake up the components? {\underline{A}}: If either situation changes, we need to wake up the component. If the component is no longer idle or the outgoing buffer on the critical path is freed up, we need to wake up the component.
\item {\underline{Q}}: What element can easily obtain the information that these two situations are being resolved? {\underline{A}}: The port knows the best. If an idle component receives a request, the component needs to wake up. And if the outgoing buffer (managed by the port) changes from full to not full, the component must wake up.
\end{itemize}

% \begin{figure}[t!]
%  \centering
%  \includegraphics[width=0.5\linewidth]
%  {images/tick_func.pdf}
%  \caption{The interface that needs to be implemented by a \texttt{TickingComponent} is an extremely simple \texttt{Tick} method, using a boolean return value that indicates if the tick has made progress. All other heavy lifting, such as stopping the ticking and waking up the component, is handled by the Akita simulator engine.}
% \label{fig:tick_func}
% \end{figure}

\begin{wrapfigure}{r}{0.5\textwidth}
\centering
\includegraphics[width=0.4\textwidth]
{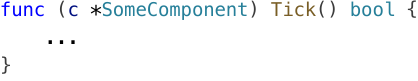}
\caption{The interface that needs to be implemented by a \texttt{TickingComponent} is an extremely simple \texttt{Tick} method, using a boolean return value that indicates if the tick has made progress. All other heavy lifting, such as stopping the ticking and waking up the component, is handled by the Akita simulator engine.}
\label{fig:tick_func}
\end{wrapfigure}

\textbf{Smart Ticking implementation.} Following the Q/A, Akita defines a special component type, Ticking Component. A Ticking Component only needs to define one method---\texttt{Tick} (see \autoref{fig:tick_func}), which returns a boolean value.
If the method returns True, the tick is making progress; and if the method returns false, the tick failed to make any progress.
The developer needs to determine whether the component is making progress in the cycle, which the program can easily determine.

Akita manages all the event scheduling behind the scenes by applying 4 simple rules. 1) When a message arrives at a component, Akita schedules a tick event in the next cycle to wake up the component. 2) When an outgoing buffer of the component changes from the full to the not-full state, Akita needs to wake up the component by scheduling a tick. 3) If a tick function returns True, meaning it is progressing, Akita schedules a tick event in the next cycle. Otherwise, Akita put the component to sleep by not scheduling a future tick event. 4) If there is already a tick event scheduled for the component, Akita will never schedule another one, preventing a component from having multiple tick events simultaneously.

% \begin{figure}[t!]
%  \centering
%  \includegraphics[width=0.8\linewidth]
%  {images/Smart_Ticking.drawio.pdf}
% \caption{Smart Ticking automatically wakes a component when a new message arrives and when a port's outgoing buffer becomes available. Smart Ticking can also put a component to sleep when the component is not making progress.}
% \label{smart_ticking}
% \end{figure}

\begin{wrapfigure}{r}{0.6\textwidth}
\centering
\includegraphics[width=0.6\textwidth]
{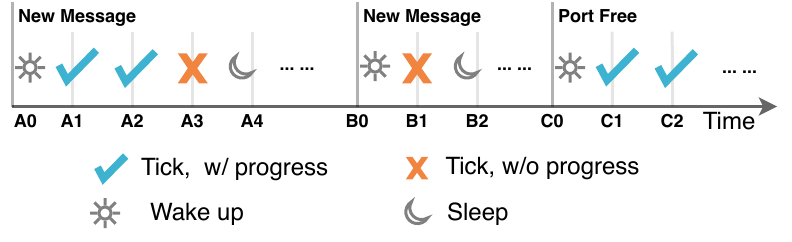}
\caption{Smart Ticking automatically wakes a component when a new message arrives and when a port's outgoing buffer becomes available. Smart Ticking can also put a component to sleep when the component is not making progress.}
\label{smart_ticking}
\end{wrapfigure}

\textbf{Smart Ticking example.}
We demonstrate how Smart Ticking works with an example, as shown in \autoref{smart_ticking}. Our component becomes active in cycle A1 after it receives a new message in A0, and the ticking progresses until it does not progress in cycle A3. After some time, when a new message arrives, Akita wakes up the component at B1. This time, the component cannot make progress, likely because it is stuck due to full outgoing buffers; the arrival of a new message does not resolve the root cause of being stuck. In this case, Smart Ticking will stop ticking. Finally, at cycle C1, once the outgoing buffer has an available slot, the component resumes ticking until the component is not making progress anymore.

As shown in the example, Smart Ticking cannot eliminate all unnecessary ticks but can significantly reduce them. An unnecessary tick is always needed to detect if the component is not progressing. Moreover, false alarms that wake up the component by mistake can happen. This is because we take a conservative approach to guarantee that the automatically skipped ticks will not impact simulation accuracy.

\textbf{Availability Backpropagation.}
Smart Ticking relies on the ports to wake up their owning components when the outgoing buffer becomes available. Availability Backpropagation expands the mechanism to also apply to connections.

% \begin{figure}[t!]
%  \centering
%  \includegraphics[width=0.6\linewidth]
%  {images/availability_backprop.drawio.pdf}
% \caption{The Availability Backpropagation mechanism. Ports and connections are not limited to sending messages forward but also propagate availability information backward, waking up proper components and connections, and improving performance. Whether to wake up the top-left or the top-right component depends on the connection arbitration policy.}
% \label{fig:availability_backprop}
% % \vspace{-0.5cm}
% \end{figure}

\begin{wrapfigure}{r}{0.5\textwidth}
\centering
\includegraphics[width=0.5\textwidth]
{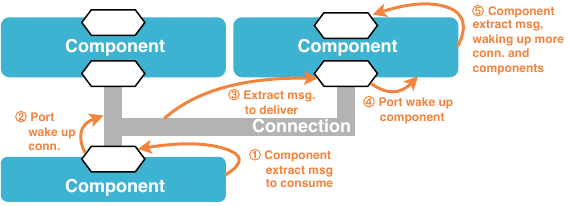}
\caption{The Availability Backpropagation mechanism. Ports and connections are not limited to sending messages forward but also propagate availability information backward, waking up proper components and connections, and improving performance. Whether to wake up the top-left or the top-right component depends on the connection arbitration policy.}
\label{fig:availability_backprop}
\vspace{-0.3cm}
\end{wrapfigure}

As shown in~\autoref{fig:availability_backprop}, the availability backpropagation process starts when a component extracts one message from one of its incoming buffers (see~\autoref{fig:availability_backprop}). If the buffer changes from the full state to the not-full state, the port wakes up the connection (see~\autoref{fig:availability_backprop}) so that the connection can deliver the new messages to the port. Note that this is different from the mechanism mentioned in Smart Ticking. Smart Ticking wakes up the component when the outgoing buffer changes state. Availability Backpropagation wakes up the connection when the incoming buffer changes state.

Then, the newly woken connection will try to deliver a message from a source port to the freed-up buffer (see~\autoref{fig:availability_backprop}), retrieving messages from the outgoing buffer of the message source port. If this source port's outgoing buffer changes from the full state to the not-full state, it wakes up the component (see~\autoref{fig:availability_backprop}), as described in the Smart Ticking mechanism. Finally, the component can process more messages, waking up more connections and upstream components (see~\autoref{fig:availability_backprop}).

\subsection{Straightforward Parallel Simulation}

Akita supports simulating hardware operations in parallel. Modern computer systems inherently embrace parallel computation, indicating that many units run concurrently. This allows us to simulate the execution of these units in parallel.

Akita employs a parallel simulation engine for parallel simulation. The parallel engine triggers events that are scheduled at the same time in parallel. We embrace a conservative parallel discrete-event simulation (PDES)~\cite{fujimoto1990parallel} method, which maintains the chronological order of the events and will not impact simulation accuracy.

The parallel simulation design is fully transparent to developers who implement models for a new hardware component. Programmers can still write code in the cycle-based simulation style, even without adding locks (satisfying \underline{DX-3}). This is possible because we forbid one component from calling any function of another component and running more than one instance of the tick function simultaneously. Akita implements most elements that may involve race conditions (e.g., ports), and properly locks the resources to prevent data races.

\begin{wraptable}{r}{0.5\textwidth}
% \vspace{-1cm}
\caption{The data traced by each task.}
\label{tab:task}
\footnotesize
\centering
\begin{tabular}{@{}llp{2.2cm}p{1.7cm}@{}}
\toprule
Field & Format & Description & Example\\
\midrule
ID & Text & The unique identifier of a task & \texttt{5C9dX8} \\
Parent ID & Text & The ID of the parent task & \texttt{7F3sY2} \\
Category & Text & High-level category & \texttt{Instruction} \\
Action & Text & The job of the task & \texttt{Mem Read} \\
Location & Text & The hardware component that carries out the task & \texttt{CPU1.Core1} \\
Start/ End & Time & The time that a hardware component starts to process/completes processing the task & 0.00014566s \\
Tags & Text List & Annotations to the task & \texttt{["cache hit"]} \\
Details & JSON & Any other information & \texttt{\{"inst":"add"\}} \\
\bottomrule
\end{tabular}
\end{wraptable}

% \takeaway{Akita's parallel simulation mechanism improves simulation performance without impacting accuracy. The parallel simulation technique used in Akita improves performance without adding any burden to developers, fulfilling \underline{UX-3} and \underline{DX-3}.}

% \takeaway{Akita's parallel simulation mechanism improves simulation performance without impacting accuracy. The parallel simulation technique used in Akita improves performance without adding any burden to developers, fulfilling \underline{UX-3} and \underline{DX-3}.}

\subsection{Tracing System} \label{sec:tracing}

Akita introduces a tracing system that enables users to extract performance metrics from the simulator, thereby facilitating a deeper understanding of the hardware behavior within a simulated system.

\textbf{Data to trace.} A fundamental design question for the tracing system is what to trace. Akita selects to trace \emph{Tasks} (see~\autoref{tab:task}) to balance generality, expressiveness, and programming overhead.

We require every task to record its parent task. For example, a memory transaction may have originated from a load instruction. In this case, the instruction task is the parent task of the memory transaction task. By recording parent tasks, we organize the tasks as a tree, creating more opportunities to analyze the data (satisfying \underline{UX-5}).

\textbf{Tracing system implementation.} The tracing system employs an Aspect-Oriented Programming (AOP) paradigm~\cite{kiczales1997aspect} to ensure flexibility and code quality. We consider one aspect to be the digital logic of the hardware and the other aspect to be the data to be collected, leading to a clear separation of concerns. Following this guidance, we develop a tracing instrumentation API and a set of tracers. Developers call the instrumentation API in the hardware-modeling code. Users can select tracers to apply to the components from which they may want to collect data.

\textbf{Instrumentation API.}
When implementing a component, the developer needs to annotate the component code using the instrumentation APIs. We design the instrumentation API to minimally interrupt the flow of the hardware logic. The instrumentation APIs provide a set of functions that developers can use to record the start/end of a task and annotate task tags (satisfying \underline{DX-5}).  More specifically, we provide three functions: i) \texttt{StartTask}, ii) \texttt{EndTask}, and iii) \texttt{TagTask} so that the developer can call them at proper times.

\textbf{Tracers.} Tracers handle the annotations generated by instrumentation API calls, deciding the action to take (e.g., record the task in the database and count tasks completed). Users can attach multiple tracers to one component and apply one tracer to multiple components. Akita provides a few first-party tracers (listed below) that can handle many common data-collection processes. The versatility of the tracers and the capability of allowing users to define their own tracers fulfill the diverse performance metrics requirement (\underline{UX-5}).

\begin{itemize}
\item \textbf{TotalTimeTracer} and \textbf{AverageTimeTracer} can calculate the total/average time of handling a certain type of task. For example, we can use the AverageTimeTracer to record the average latency of memory transactions at a certain level of cache.

\item \textbf{BusyTimeTracer} calculates the time that a certain component is at least handling one type of task. For example, it can be used to provide data for ALU utilization (the percentage of time that the ALU is busy).

\item \textbf{TagCountTracer} can count the occurrences of specific tags, such as cache hits and misses.

\item \textbf{DBTracer} stores all the tasks in a database. Akita supports SQLite (default), MySQL, MongoDB, and Comma-Separated Values (CSV) files. The recorded tasks form a full execution trace and can be used to analyze hardware behavior after simulation.
\end{itemize}

% \begin{figure}[t!]
%  \centering
%  \includegraphics[width=0.6\linewidth]
%  {images/backtrace_v2.drawio.pdf}
% \caption{Comparing the backtrace provided by the programming language (a) and Akita (b), Akita's backtrace maintains more architecture-related contextual information.}
% \label{fig:backtrace}
% % \vspace{-0.5cm}
% \end{figure}

\textbf{Enhanced backtracing.} Although the tracing system design is straightforward, it can support many features not commonly available in other simulators. Enhanced backtracking is an example.

\begin{wrapfigure}{r}{0.6\textwidth}
\centering
\includegraphics[width=0.6\textwidth]
{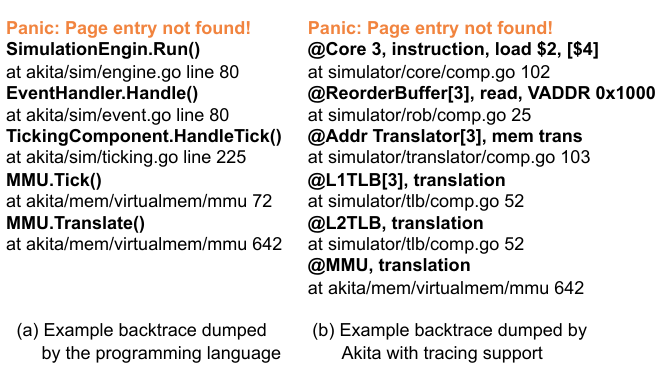}
\caption{Comparing the backtrace provided by the programming language (a) and Akita (b), Akita's backtrace maintains more architecture-related contextual information.}
\label{fig:backtrace}
\vspace{-0.3cm}
\end{wrapfigure}

When developing event-driven simulation, a problem that many developers face is that the function-based backtrace can provide limited information (see~\autoref{fig:backtrace} (a)). This is because event-driven simulators handle progress event by event, and the only contextual information available is the current event.
Akita addresses this issue by utilizing its access to tree-based task structures. In case of a program crash (or when a developer needs to print a backtrace), Akita only requires the developer to provide the current task that triggered the crash. Relying on the parent task field to trace back to the root task, Akita will capture more informative backtraces (see~\autoref{fig:backtrace} (b)). The Akita backtraces, printed alongside the programming language traces to the terminal, can better support debugging, fulfilling \underline{DX-4}.

\textbf{Performance analysis framework.} Akita provides an engineered solution to record performance data for later analysis. For each simulation, we record all the performance-related data in an SQLite database record, which can include multiple tables (satisfying \underline{UX-5}). We dedicate one table to record execution information (e.g., terminal command, start/end time, working directory) and whole simulation performance metrics (e.g., the number of instructions executed).

Besides the simulation-level performance metrics, Akita records data periodically from ports and buffers. Akita provides specialized tracers to periodically record buffer levels and each port's input/output throughput.  We leave interfaces for developers to add additional types of data to collect.
We focus on buffer levels and port throughput because they can help identify performance bottlenecks. For example, if a port is delivering data at a constant rate for a long time, it suggests that the component delivering the data is working at full speed, and it is likely to be a performance bottleneck. Additionally, if a buffer level remains high for a long time, the component consuming the message from the buffer is likely not fast enough.

% To fulfill the requirement of examining performance while the simulation is running, Akita allows simulator developers to turn a running simulation into a web server and provides a web portal to track buffer levels and achieve bandwidth in real time, fulfilling \underline{UX-4c}.

% \textbf{Real-time performance monitoring with AkitaRTM~\cite{mosallaei2023looking}.} Akita can help developers check how the system under simulation performs while the simulation is running through a real-time monitoring tool. Besides, Akita can effectively identify both deadlocks and livelocks by using its comprehensive monitoring and debugging tool. The integrated Bottleneck Analyzer identifies when components are stuck due to deadlocks, as seen with buffer content mismatches, and also helps detect livelocks by monitoring patterns of repetitive, non-productive activities within the simulation, fulfilling \underline{UX-4c} and \underline{DX-4b}.

\subsection{AkitaRTM}
Akita provides developers with real-time performance insights through its integration with AkitaRTM~\cite{mosallaei2023looking}, enabling users to monitor how the system under simulation performs when the simulation is running. AkitaRTM can quickly identify problems with running simulations, perform ad-hoc analysis, and debug simulator hangs, fulfilling \underline{UX-4}. When any Akita-based simulation starts, Akita spawns a server that servers the AkitaRTM website and providing APIs that allow inspection and controlling.

\begin{wrapfigure}{r}{0.7\textwidth}
\centering
\includegraphics[width=\linewidth]{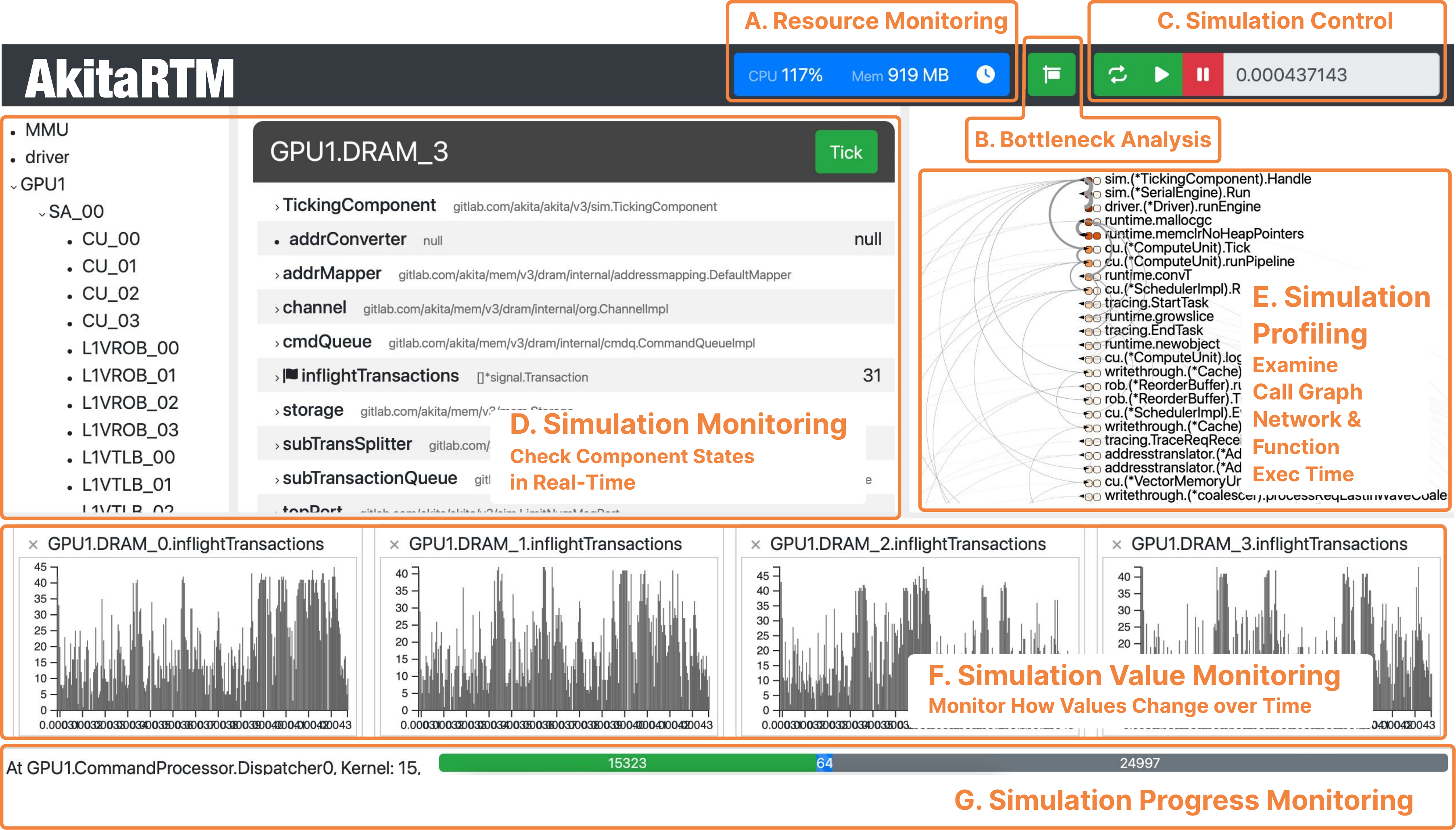}
\caption{AkitaRTM provides real-time visibility into simulator execution, including resource utilization, simulation progress, component states, bottleneck analysis, and profiling information, enabling interactive monitoring and debugging of running Akita-based simulations.}
\label{fig:akitaRTM}
\vspace{-0.5cm}
\end{wrapfigure}

The most important part of Akita is the simutation monitoring feature, which allows a user to check all the components and being simulated and examine the value of the fields of the components in realtime (see \autoref{fig:akitaRTM} (D)). Additionally, if user select a field with the flag button, AkitaRTM shows a realtime visualization at the bottom. Other than monitoring, AkitaRTM also presents simulation progress (estimated), pprof-based~\cite{pprof} simulation profiling, and buffer-level-based bottleneck analysis.

Akita streamlines the debugging process by helping developers identify and resolve simulation hangs, ensuring accurate performance evaluation. The process typically begins with real-time simulation monitoring. A hang can be detected through several indicators: lack of forward progress in simulation, CPU usage dropping well below 100\%, and persistent buffer contents suggesting stalled components.

Once a hang is detected, developers use AkitaRTM's Bottleneck Analyzer to locate the stalled components. In a successful simulation, all buffers should be empty upon completion; a non-empty buffer indicates that its owning component is unable to proceed. With AkitaRTM, there is no need to restart the simulation or set breakpoints from scratch to diagnose the issue. Instead,developers can set a breakpoint at the first line of the \texttt{Tick} function of a suspect component. AkitaRTM can force-trigger a tick on that component, causing the breakpoint to fire. From there, the developer can step through the code line by line to pinpoint exactly why the component cannot make progress (satisfying \underline{UX-4}).

% By providing a structured debugging approach, Akita enables developers to systematically trace and resolve performance bottlenecks with minimal interruption in the simulation workflow. This reduces the debugging time and enhances efficiency, particularly in large-scale or long-running simulations. By integrating real-time monitoring with debugging, Akita provides support to developers in diagnosing simulation errors efficiently.

\subsection{Visualization with Daisen}

\begin{figure}
\centering
\includegraphics[width=\textwidth]{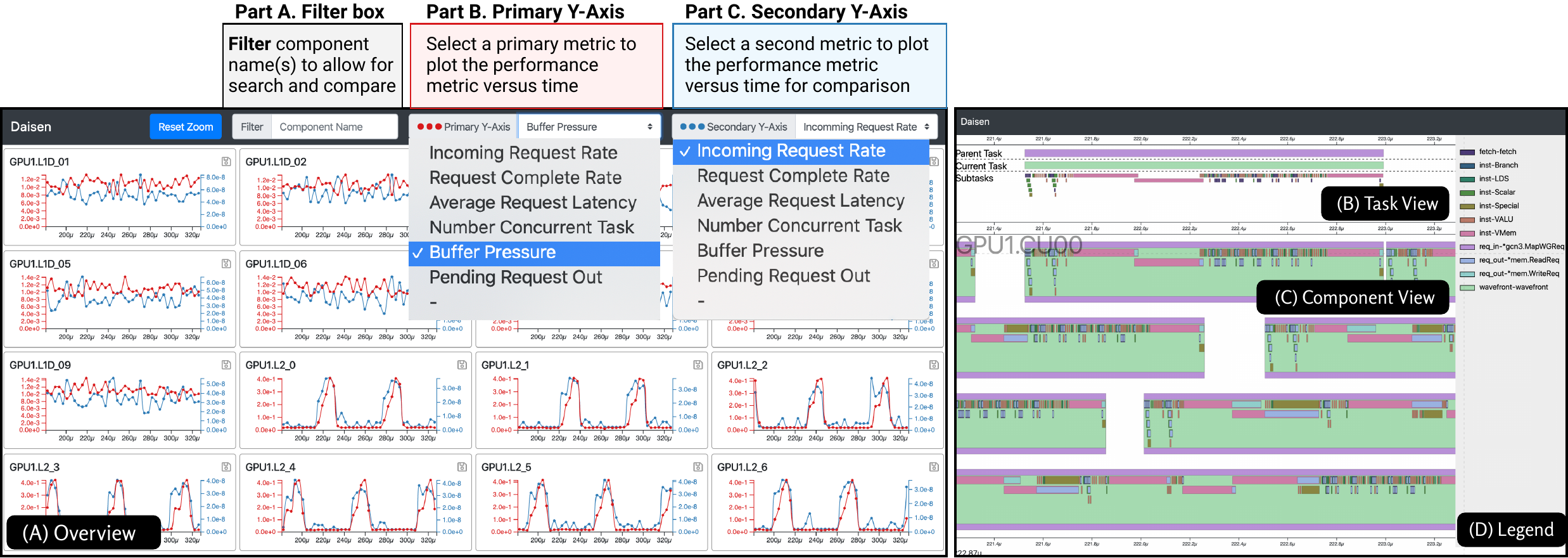}
\caption{Daisen visualization integrated with Akita, showing (A) an overview of component-level performance metrics over time, (B) hierarchical task relationships, (C) per-component task execution timelines, and (D) a task legend. This interface enables interactive inspection of simulator behavior without modifying simulator code.}
\label{fig:daisen}
\vspace{-0.5cm}
\end{figure}

Beyond real-time debugging, understanding simulator behavior also requires examining what happened during a simulation after it completes. Akita records all completed tasks (e.g., instructions, cache transactions, and inter-component messages) throughout a simulation, and includes Daisen~\cite{sun2021daisen}, a web-based visualization tool, as its built-in trace visualization subsystem (see~\autoref{fig:daisen}). Any simulator built with Akita can be visualized out of the box, if the components are instrumented properly.

Daisen provides analysis at multiple levels of abstraction. Its Overview Panel displays time-aligned performance metrics for all hardware components, enabling users to quickly spot execution phases, anomalies, or stalled progress. Users can then drill down into individual components to inspect every executed task, or explore hierarchical Task Views that reveal how high-level activities decompose into lower-level component interactions.

This tight integration between Akita's simulation engine and Daisen's visualization allows users to move seamlessly from a global overview to individual component execution without adding too much logging or debugging code, making complex simulator behavior transparent and explainable (fulfilling \underline{UX-5}).

\section{Evaluation}

We evaluate Akita using both technical- and human-centric methods. In this section, we demonstrate Akita's technical capability.

\begin{wraptable}[16]{r}{0.4\textwidth}
  \vspace{-1cm}
  \caption{Benchmarks used in the experiments.}
  \label{tab:benchmarks}
  \footnotesize
  \centering
  \begin{tabular}{@{}llp{2.6cm}@{}}
    \toprule
    Abbr. & Suite & Workload \\
    \midrule
    AES  & HeteroMark & Advanced Encryption Standard \\
    ATAX & PolyBench  & Matrix Transpose and Vector Multiplication \\
    FFT  & SHOC       & Fast Fourier Transform \\
    FIR  & HeteroMark & Finite Impulse Response Filter \\
    FW   & AMDAPPSDK  & Floyd--Warshall Algorithm \\
    KM   & HeteroMark & KMeans Clustering \\
    MM   & AMDAPPSDK  & Matrix Multiplication \\
    MT   & AMDAPPSDK  & Matrix Transpose \\
    ReLU & DNNMark    & Rectified Linear Unit \\
    SC   & AMDAPPSDK  & Simple Convolution \\
    S2D  & SHOC       & Stencil 2D \\
    \bottomrule
  \end{tabular}
\end{wraptable}

% \begin{table}[tb]
%   \caption{Benchmarks used in the experiments. }
%   \label{tab:benchmarks}
%   \footnotesize%
%   \centering%
%   \begin{tabular}{@{}llp{5.0cm}@{}}
%   \toprule
%   Abbr. & Suite & Workload \\
%   \midrule
%   AES & HeteroMark & Advanced Encryption Standard \\
%   ATAX & PolyBench & Matrix Transpose and Vector Multiplication\\
%   FFT & SHOC & Fast Fourier Transform\\
%   FIR & HeteroMark & Finite Impulse Response Filter\\
%   FW & AMDAPPSDK & Floyd-Warshall Algorithm\\
%   KM & HeteroMark & KMeans Clustering\\
%   MM & AMDAPPSDK & Matrix Multiplication \\
%   MT & AMDAPPSDK & Matrix Transpose\\
%   ReLU & DNNMark & Rectified Linear Unit\\
%   SC & AMDAPPSDK & Simple Convolution\\
%   S2D & SHOC & Stencil 2D\\
%   \bottomrule
%   % \vspace{-0.8cm}
%   \end{tabular}%
% \end{table}

\subsection{Methodology}

\textbf{Platform.} Our experiments are conducted on an AMD EPYC 7302P 16-core processor, 128GB memory, and Ubuntu 20.04.6 LTS operating systems. All programs are compiled with Go 1.23.0. We use the \texttt{GOMAXPROCS} environment variable to control the number of cores used in parallel simulation.

\textbf{Simulation configuration.}MGPUSim\cite{sun2019mgpusim} simulator was developed with Akita simulation engine. We evaluate MGPUSim performance by toggling on/off of some Akita features.
%Since MGPUSim is also written in Go, we can easily migrate MGPUSim to Akita.

\textbf{Benchmarks.} To ensure the robustness of our evaluation, we utilized a wide range of representative benchmarks (see~\autoref{tab:benchmarks}) from many benchmark suites, including AMDAPPSDK~\cite{AMDAPPSDK}, SHOC~\cite{danalis2010scalable}, HeteroMark~\cite{sun2016hetero}, PolyBench~\cite{polybench}, and DNNMark~\cite{dong2017dnnmark} (see \autoref{tab:benchmarks}). These suites cover a majority of the workloads MGPUSim supports natively. We run large problem sizes in order to limit the impact of workload startup.  We simulate an AMD R9 Nano GPU configuration, which should have a limited impact on the generalizability of the evaluation presented.

\subsection{Smart Ticking Performance}

\begin{figure}[t]
  \centering
  % Row 2 (centered)
  \begin{subfigure}{0.45\linewidth}
    \centering
    \includegraphics[width=\linewidth]{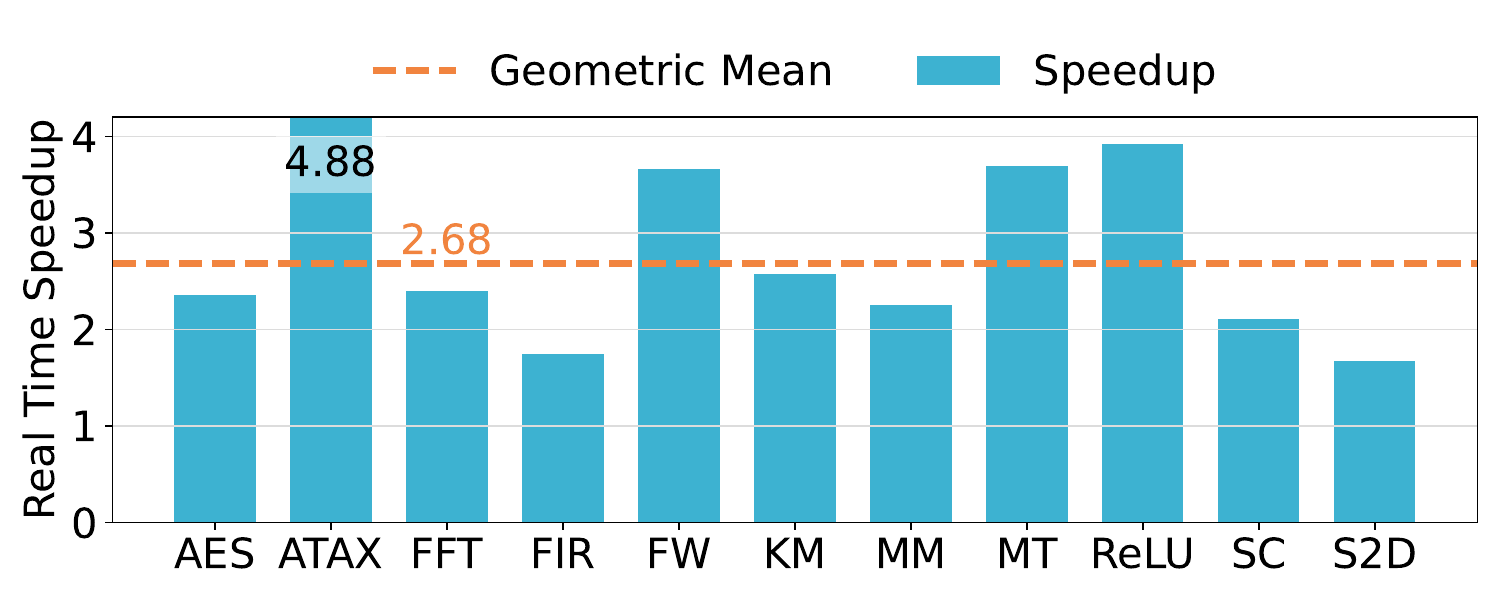}
    \caption{Smart Ticking speeds up simulations by 2.68X on average.}
    \label{fig:smart_ticking_speedup}
  \end{subfigure} \hspace{0.02\linewidth}
  \begin{subfigure}{0.46\linewidth}
    \includegraphics[width=\linewidth]{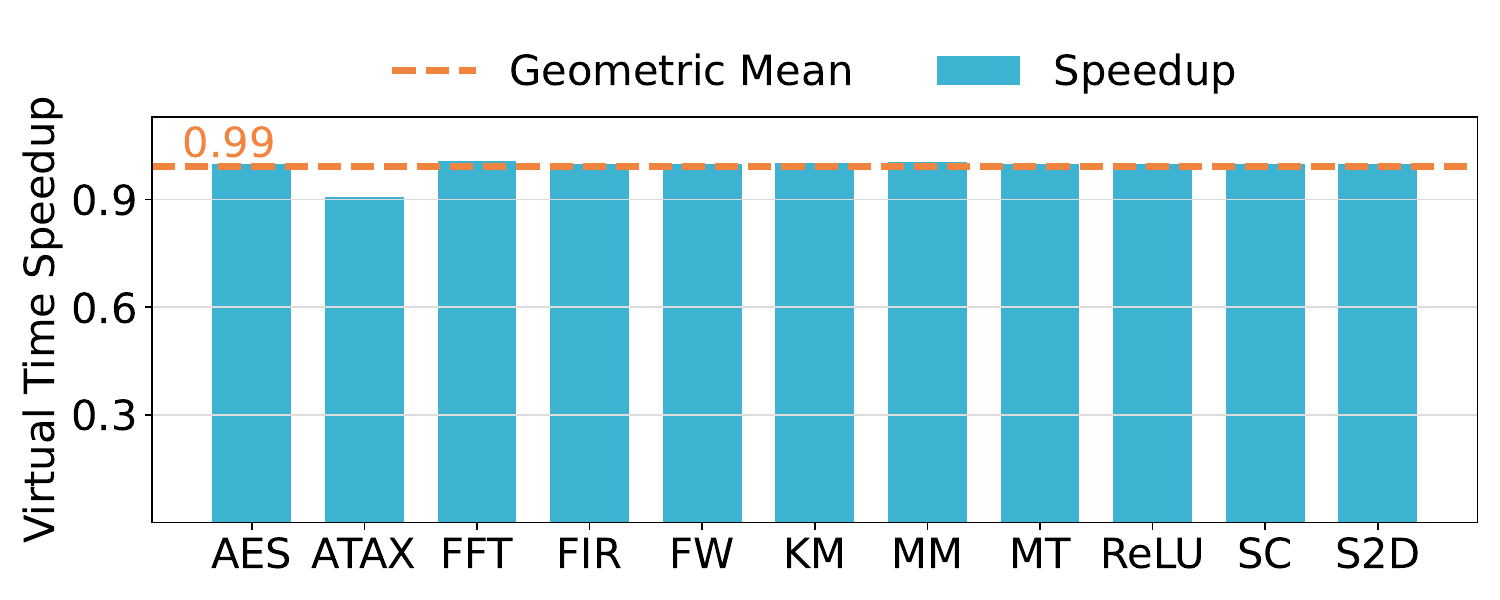}
    \caption{Smart Ticking introduce less than 1\% of simulation error.}
    \label{fig:smart_ticking_speedup_accuracy}
  \end{subfigure}

  \caption{The impact of Smart Ticking on real time (i.e., simulation execution time) and virtual time (i.e., the estimated execution time of the simulated workload).}
  \label{fig:combined_smart_ticking}
\end{figure}

We compare two versions of MGPUSim, one with Smart Ticking and one without. We focus on the execution time of the simulation when these two versions of MGPUSim run the same workload. As shown in \autoref{fig:smart_ticking_speedup}, Smart Ticking improves performance by 2.68X on average. Some benchmarks, such as ATAX, demonstrate high speedup. ATAX has limited parallelism and cannot utilize all the cores, so Smart Ticking puts those idle cores to sleep. Additionally, as we validate the accuracy of Akita against MGPUSim (see~\autoref{fig:smart_ticking_speedup_accuracy}), we observe that Akita does not introduce noticeable errors in most of the benchmarks. Smart ticking introduces less than 1\% of error to an existing simulator.

\begin{wrapfigure}{r}{0.58\textwidth}
  \centering
  \includegraphics[width=0.6\textwidth]{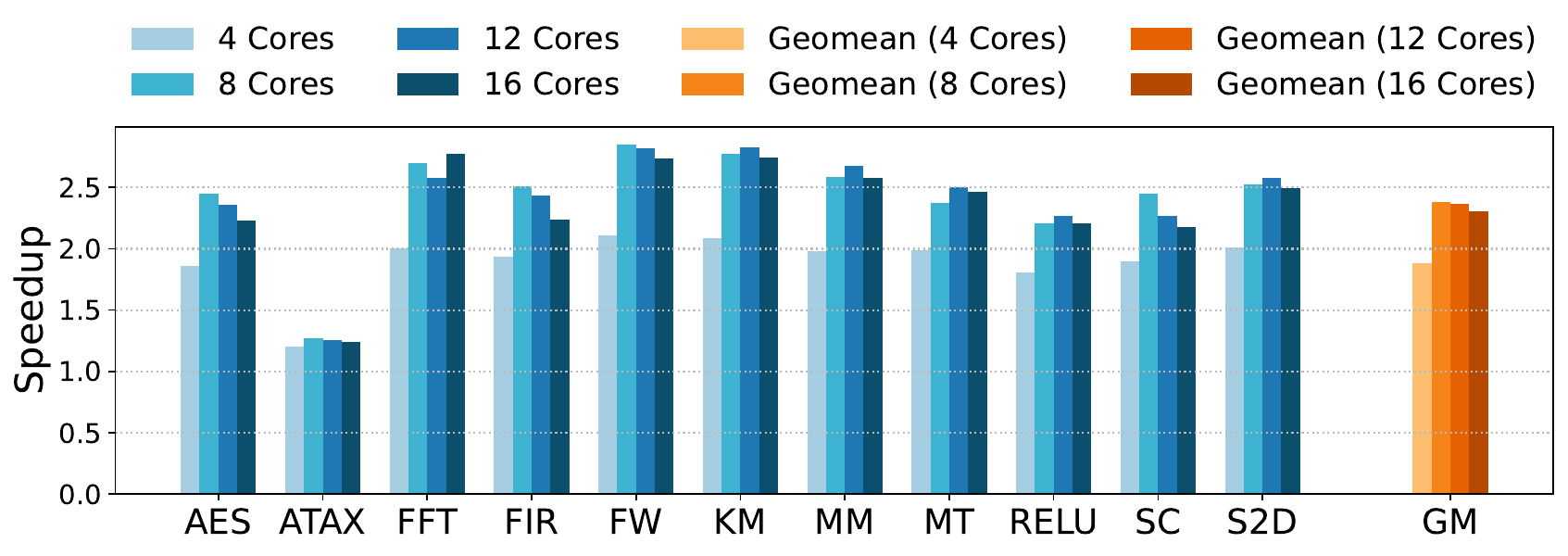}
  \caption{The speedup achieved through parallel implementation with detailed timing simulation across various benchmarks, demonstrating notable performance improvement.}
  \label{fig:parallel}
\end{wrapfigure}

\subsection{Parallel Simulation Performance} \label{sec:parallel_perf}

% Next, we apply the Akita parallel simulation engine to MGPUSim to evaluate the performance improvement. We run a set of benchmarks in both emulation (emulate instructions without modeling the memory system) and timing simulation mode.

% \begin{wrapfigure}{r}{0.58\textwidth}
%  \centering
%  \includegraphics[width=0.6\textwidth]{images/parallel_v3.pdf}
% \caption{The speedup achieved through parallel implementation with detailed timing simulation across various benchmarks, demonstrating notable performance improvement. \yifan{More cores}}
% \label{fig:parallel}
% \end{wrapfigure}

Next, we apply the Akita parallel simulation engine to MGPUSim to evaluate the performance improvement. Specifically, we have applied the Akita parallel simulation engine to MGPUSim using configurations of 4 cores, 8 cores, 12 cores, and 16 cores. Note that, as we run the parallel simulation with Smart Ticking turned on, this speedup is an addition to Smart Ticking speedup.

Our evaluation shows that Akita's parallel simulation methods improve the performance by 1.88X, 2.38X, 2.37X and 2.3X for 4 cores, 8 cores, 12 cores, and 16 cores respectively (see ~\autoref{fig:parallel}). The performance tops when about 8 cores are used, effectively reducing a simulation that would run more than 2 days to a 24-hour time frame. ATAX does not see performance improvement mainly because it does not have sufficient parallelism, which is a situation mainly addressed by Smart Ticking.

\subsection{Overhead of tracing system.}

Next, we evaluate the performance overhead of the tracing system. We add performance metrics tracers for the components in MGPUSim to reflect a typical use case. More specifically, we add the following tracers, with the leading numbers representing the number of tracers of the type applied.

\begin{wrapfigure}{r}{0.5\textwidth}
  \centering
  \includegraphics[width=0.5\textwidth]{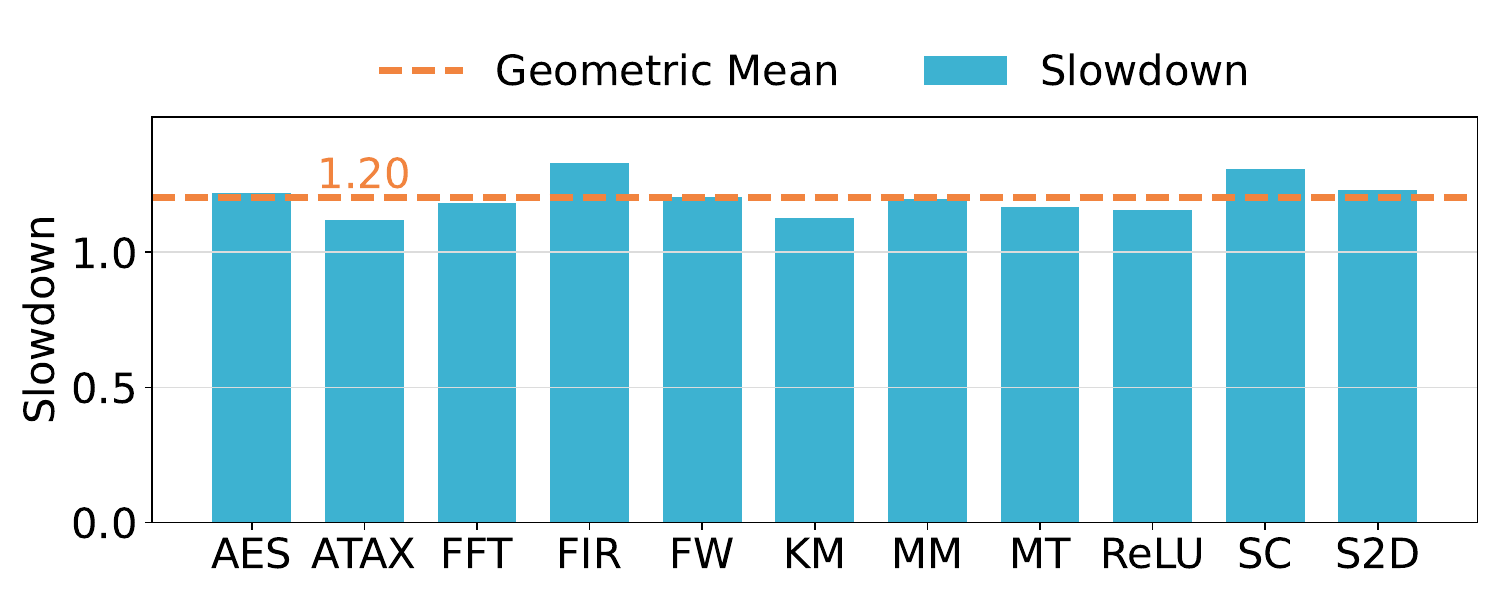}
  \caption{The simulation slowdown due to the inclusion of tracers to capture performance metrics.}
  % \vspace{-0.5em}
  \label{fig:report_all}
\end{wrapfigure}

\begin{itemize}
  \item [\textbf{64}] An instruction count tracer for each type of instruction executed in each core.
  \item [\textbf{64}] A customized tracer for the Cycle-Per-Instruction Stack~\cite{harris2010digital} for each core.
  \item [\textbf{256}] A busy time tracer for the ALU in each subcore (4 subcore per core).
  \item [\textbf{112}] An average request latency tracer for each L1 and L2 cache.
  \item [\textbf{209}] A cache hit rate tracer for each L1 cache, L2 cache, L1 TLB, and L2 TLB.
  \item [\textbf{32}] A transaction count for each DRAM controller.
  \item [\textbf{32}] An average read/write latency for each DRAM controller.
\end{itemize}

Overall, the tracers create a 20\% slowdown on average (see~\autoref{fig:report_all}). The slowdown is relatively consistent throughout the benchmarks tested. Considering the rich performance metrics collected and reported, the slowdown is acceptable. The non-negligible slowdown also supports our design decision not to always collect all the metrics at all times and allows users to select the most useful metrics to collect and avoid unnecessary slowdown.

\section{Case Studies}

Next, we perform two case studies to demonstrate the versatility of Akita.

\subsection{Onira}

To demonstrate the generality of Akita, we build an in-order RISC-V timing model using the Akita engine. The simulator mirrors a standard five-stage pipeline architecture, supporting hazard prevention, value forwarding, and a memory interface. Developing the timing model took a master's student with minimal computer architecture experience (only one graduate-level course in the last author's class) 2--3 weeks, including time to learn Go and the Akita engine.

\begin{wrapfigure}{r}{0.8\textwidth}
  \centering
  \includegraphics[width=0.8\linewidth]{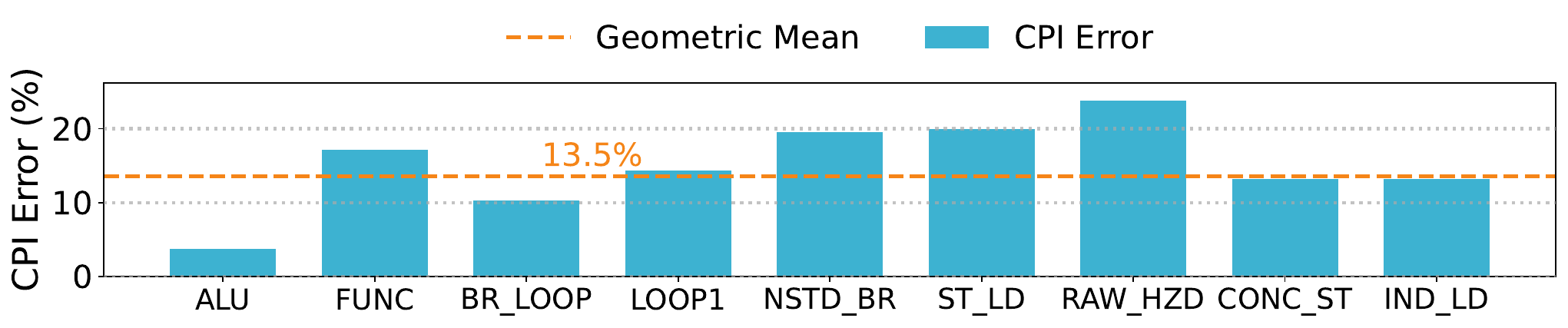}
  \caption{CPI error between Onira and RTL across selected microbenchmarks and concurrent tests.}
  \label{fig:final_validation}
\end{wrapfigure}

Accuracy is measured by comparing the Akita-based model against a Verilator RTL implementation under identical conditions (single core, 5-cycle memory latency).
We use hand-written microbenchmarks that isolate specific behaviors: ALU execution (ALU), function calls (FUNC), loop and branch patterns (BR\_LOOP, LOOP1, NESTED\_BR), a store--load pair (ST\_LD), and a RAW hazard case (RAW\_HZD). The concurrent tests issue multiple independent memory operations (CONC\_ST, IND\_LD). For each benchmark, we record the number of instructions, cycles, and CPI on both simulators.

Across these workloads, the Akita-based model stays within roughly 10--20\% CPI error relative to RTL, with most compute- and control-oriented tests below 15\% (see \autoref{fig:final_validation}). Larger deviations appear in load-use and write-heavy patterns, which are more sensitive to microarchitectural timing.

% Our validation uses a suite of hand-written microbenchmarks and concurrent tests that each isolate a specific architectural behavior. The microbenchmarks include straight-line ALU code (ALU), function calls (FUNC), loop and control-flow patterns (BR\_LOOP, LOOP1, NESTED\_BR), a simple store-then-load (ST\_LD), and a RAW data-hazard case (RAW\_HZD). The concurrent tests issue multiple independent stores and loads in parallel (CONC\_ST, IND\_LD). For each benchmark, we collect total instructions and cycles and compute CPI on both the Akita-based model and the RTL simulator.

% \begin{figure}[t!]
%  \centering
%  \includegraphics[width=0.8\linewidth]{images/final_validation_plot.pdf}
% \caption{CPI error between RISCV CPU Simulator and RTL across selected microbenchmarks and concurrent tests.}
% \label{fig:final_validation}
% \end{figure}

% \begin{wrapfigure}{r}{0.5\textwidth}
%  \centering
%  \includegraphics[width=0.5\textwidth]{images/tracer_slowdown.pdf}
% \caption{The simulation slowdown due to the inclusion of tracers to capture performance metrics.}
% % \vspace{-0.5em}
% \label{fig:report_all}
% \end{wrapfigure}

\begin{figure}[t!]
  \centering

  \begin{subfigure}[b]{0.3\linewidth}
    \centering
    \includegraphics[width=\linewidth]{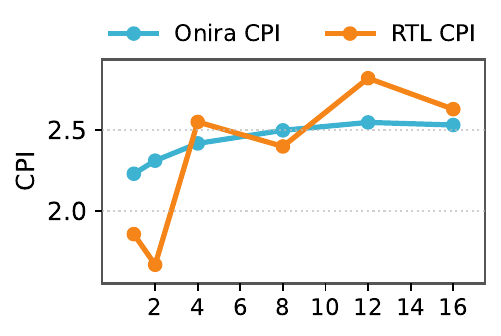}
    \caption{MLP trend comparison.}
    \label{fig:mlp_trend}
  \end{subfigure}\hspace{0.02\linewidth}
  \begin{subfigure}[b]{0.3\linewidth}
    \centering
    \includegraphics[width=\linewidth]{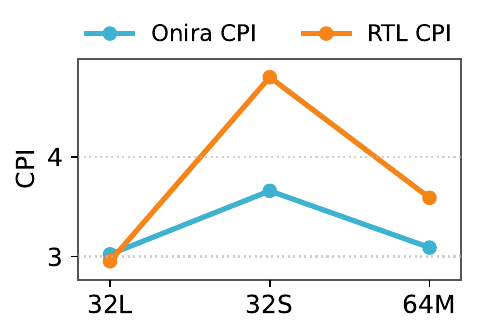}
    \caption{Burst pattern comparison.}
    \label{fig:burst_trend}
  \end{subfigure}

  \caption{Comparison of Onira vs. RTL memory behavior across different access patterns.}
  \label{fig:mlp_burst_combined}
\end{figure}

To study memory scaling, we use a memory-level parallelism (MLP) benchmark that issues $N$ independent loads ($N=1$--16). CPI increases and saturates similarly on both simulators (\autoref{fig:mlp_trend}), indicating that Akita captures queueing and bandwidth effects. Burst-style traffic (\autoref{fig:burst_trend}) shows matching qualitative behavior: store bursts are slowest, followed by mixed bursts, then load bursts, with CPI values in the expected memory-bound range.

Overall, the Akita-based timing model preserves RTL trends for pipeline and memory-intensive behaviors while providing far faster simulation, making it a practical tool for architectural exploration and instructional use, and highlighting memory-sensitive cases as targets for future refinement.

\subsection{TrioSim}

% \begin{figure}[t!]
%  \centering
%  \includegraphics[width=0.6\linewidth]{images/akita_triosim.pdf}
% \caption{Validating the trace-based simulation against a 4-NVIDIA A40 GPU system running PyTorch training workloads with different parallelism configurations. DP=data parallelism, TP=tensor parallelism, PP=Pipeline parallelism. }
% \label{fig:triosim_validation}
% \end{figure}

\begin{wrapfigure}{r}{0.5\textwidth}
  \centering
  \includegraphics[width=0.5\textwidth]{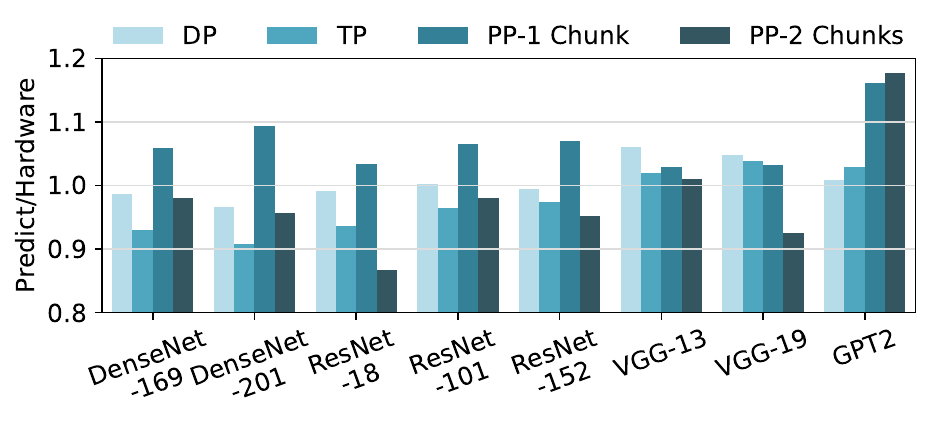}
  \caption{Validating the trace-based simulation against a 4-NVIDIA A40 GPU system running PyTorch training workloads with different parallelism configurations. DP=data parallelism, TP=tensor parallelism, PP=Pipeline parallelism.}
  % \vspace{-0.5em}
  \label{fig:triosim_validation}
\end{wrapfigure}

Prior examples demonstrate that Akita can support a range of cycle-level simulators. Here, we use TrioSim to demonstrate that Akita can also be used to develop a high-level, trace-driven, purely event-driven simulator.
TrioSim~\cite{li2025triosim} is a lightweight simulator for large-scale deep neural network (DNN) workloads on multi-GPU systems. The simulator consumes operator-level traces collected from single-GPU executions and extrapolates them into multi-GPU communication and computation events.

TrioSim is written in Go and built directly on Akita's event-driven framework. Using a performance model~\cite{li2023path} to estimate kernel or operator execution time, TrioSim condenses each kernel/operator into a single event and fast-forwards without simulating microarchitectural details. For data movement, TrioSim uses Akita's port--connection abstraction. To achieve higher performance and avoid cycle-by-cycle simulation of data transfer, TrioSim provides an alternative implementation of ports and connections. This implementation uses a flow-based network model~\cite{giuli2002narses} and event-driven scheduling to accelerate simulation, demonstrating Akita's adaptability.

A validation against a platform with four A40 GPUs running PyTorch DNN training workloads (see~\autoref{fig:triosim_validation}) shows that the Akita-based trace-driven DNN simulation achieves high accuracy with limited development effort.

\section{User Survey Study}

\textbf{Method.} We conduct a preliminary study to evaluate Akita's UX and DX through an IRB-approved survey of researchers who actively use Akita in their work. The survey yielded 16 responses, none from authors of this paper.

A randomized controlled study comparing Akita against another engine would require participants to be proficient in multiple simulation frameworks. Since gaining proficiency in a simulation engine typically takes days to weeks, such a study design is not practical. Instead, we survey experienced Akita users and focus on identifying Akita's relative strengths and weaknesses across different aspects of UX and DX. This within-subject design allows us to pinpoint which areas users find most and least satisfactory, providing actionable guidance for future improvements. We acknowledge that this design cannot establish absolute comparisons with other engines.

\textbf{Demographics.} The 16 participants include PhD students and researchers at varying stages of experience with Akita: 6 with 0--1 year, 6 with 2--3 years, 2 with 4--5 years, 1 with 6--8 years, and 1 with over 8 years.

\begin{figure*}[t!]
  \centering
  \includegraphics[width=.8\linewidth]{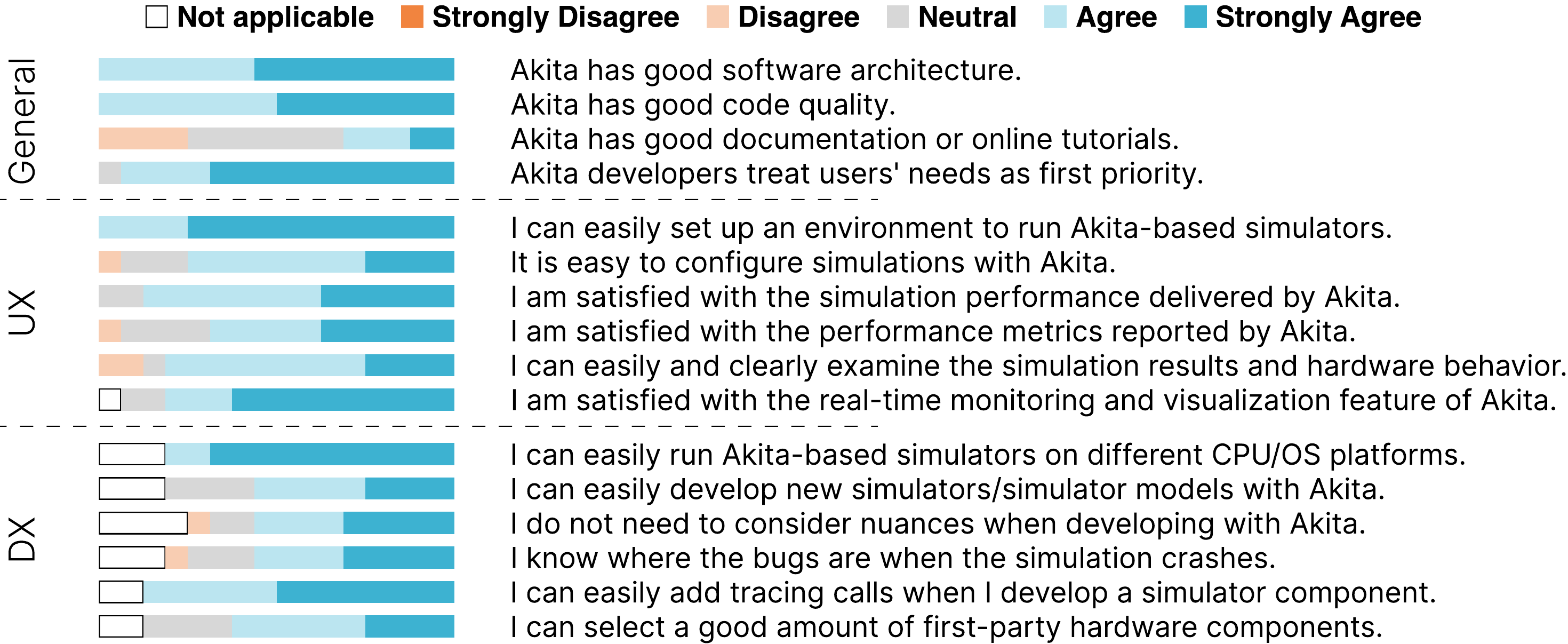}
  \caption{To what degree do you agree with the statements?}
  \label{fig:features}
\end{figure*}

\textbf{General impression.} The first section assesses users' general impressions of Akita's software architecture, code quality, documentation, and responsiveness to user needs. As shown in~\autoref{fig:features}, all 16 participants agree that Akita provides a coherent software architecture, supports the development of high-quality code, and that its developers consider user needs. However, participants rated documentation and tutorials notably lower, highlighting an area for future improvement. This result also serves as a validity check, indicating that participants were willing to report negative experiences.

\textbf{User friendliness.} The next section evaluates the user-friendliness of common tasks: setting up environments, configuring simulations, running experiments, and analyzing results. Participants rated environment setup, simulation performance, and monitoring and visualization features favorably. The areas with the lowest satisfaction were configuring new systems, the richness of available performance metrics, and support for understanding detailed hardware behavior (see~\autoref{fig:features}).

\textbf{Developer friendliness.} This section assesses Akita's support for developers, covering cross-platform deployment, component development, automation of complex concerns (e.g., smart-ticking and parallel simulation), debugging support, and the availability of first-party hardware components. As shown in~\autoref{fig:features}, cross-platform deployment and component development received the highest ratings. Automation of smart-ticking and parallel simulation, as well as debugging support, were also rated positively, though with more variance among participants.

\section{Discussion}

Akita is designed as a simulator engine rather than a simulator. In this section, we discuss the implications of this critical decision.

\subsection{Pros and Cons of Dedicated Engine}

\textbf{Pros.} 
Pros include: (1) \underline{Avoid reinventing the wheel.} Simulators often need similar optimizations. Sharing the engine allows optimization (e.g., parallel simulation, visualization support) to be implemented at the engine level and automatically applies to all the simulators.  (2) \underline{Separate of concern.} Isolating the engine allows developers to focus on the common requirements of simulators (e.g., user experience, performance, traceability). The requirement to implement a certain hardware component will not impact the simulator architecture design. (3) \underline{Uniformity.} All the components must be implemented following the same engine. The uniformity allows developers to easily understand others' code and integrate components, reducing development difficulty.

% This approach allows users and developers to easily modify and extend the simulator. Users can tailor the simulation to meet specific requirements altering the core simulation engine, and developers can integrate new components without disrupting the entire system. In addition, the system can be more adaptable to different software environments. Moreover, features such as parallel simulation and elimination of constant component polling inherently leverages this separation, allowing the engine to manage resource allocation and synchronization more effectively, while accelerating the overall simulation speed (see~\ref{sec:goals} for more details). 

\textbf{Cons.} The benefit of the engine does not come without costs, including: (1) \underline{Difficult in architecture design.} A good engine requires a careful design that considers all the needs of individual simulators and robust software architecture. We are happy to make an effort to work with the community and iteratively improve the engine by listening to the needs of researchers.  (2) \underline{More code.} A loosely coupled simulator infrastructure may require more code that defines interfaces and implements communication protocols. (3) \underline{Maintanence cost.} Once the engine bumps to a major version that cannot ensure backward compatibility, it requires all the tools to migrate to enjoy the new features. We will consider these cons and mitigate the risk while maintaining the infrastructure.

\subsection{Traditional Criteria that Does not Apply} \label{sec:validation_accuracy}
Since Akita is a simulator engine, some traditional requirements for a simulator do not apply anymore. For example, it is unfair to ask whether a simulator engine is accurate. Ensuring accuracy is the responsibility of individual simulators or each individual component. Moreover, while the engine can support the development of a wide range of simulators, including full-system or heterogeneous-system simulators, the simulator engine itself cannot simulate such systems. Therefore, a quick answer to questions like if Akita can simulate RISC-V CPUs or heterogeneous systems is no.

\section{Related Work}\label{sec:related_work}

% \subsection{Comparison with Existing Infrastructures} \label{sec:comparison_existing}

The development of a simulator usually requires considerable effort. They serve as essential infrastructures for the whole computer architecture community. Consequently, there has been a plethora of simulators for CPUs, GPUs, memory systems, networks, and domain-specific accelerators.

A large number of CPU simulators are widely available~\cite{akram201686}, including gem5~\cite{binkert2011gem5, lowe2020gem5}, SST~\cite{rodrigues2011structural} and Multi2Sim~\cite{ubal2007multi2sim}. 
A major effort has been spent on enhancing the performance of CPU simulators using either parallel simulation~\cite{sanchez2013zsim} or sampling-based simulation~\cite{carlson2011sniper}. Simulators such as gem5~\cite{binkert2011gem5}, MARSSx86~\cite{patel2011marss}, and SimFlex~\cite{wenisch2006simflex} can perform full-system simulation that simulates both the operating system and the applications.

GPU simulators focus on capturing the complexities of parallel processing and memory hierarchies. Notable GPU simulators include GPGPUSim~\cite{bakhoda2009analyzing}, Multi2Sim~\cite{ubal2012multi2sim, gong2017multi2sim}, 
gem5 AMD APU Model~\cite{gutierrez2018lost},
MGPUSim~\cite{sun2019mgpusim}, AccelSim~\cite{khairy2020accel}, 
NVArchSim~\cite{villa2021need}, and NaviSim~\cite{bao2022navisim}.

Simulators extend to more specific components. For example, Ramulator~\cite{kim2015ramulator, luo2023ramulator} and DRAMSim~\cite{wang2005dramsim, rosenfeld2011dramsim2, li2020dramsim3} are dedicated to DRAMs. Network-on-chip systems also have dedicated simulators such as DARSim~\cite{lis2010darsim}, OMNeT++~\cite{ben2011nocs}, and BookSim~\cite{jiang2010booksim}. The recent development in domain-specific accelerators also drives the development of their simulators, such as Aladdin~\cite{shao2014aladdin} and Stonne~\cite{munoz2021stonne}. 

% \subsection{Comparison with Asim and HAsim} \label{sec:comparison_asim}
% ASim/HASim is xxxx. They fulfilly the similar requirement by xxx. But they are different in xxx way.

On the same line with Akita's Availability Backpropagation feature, HAsim introduces credits~\cite{pellauer2011hasim}. The credits, working in a similar way to the credit-based network flow control mechanism~\cite{coenen2006buffer}, are sent by the receiver to the sender to notify if the sender can run the simulation for the current cycle.  HAsim’s credit feature has similarities with Akita’s Availability Backpropagation as they both allow the downstream to notify if the upstream component can make forward progress. Still, there are a few fundamental differences rooted in their use-case differences (FPGA-based vs. software-based simulation). (1) Akita’s backpropagation does not require explicit tracking. Akita only wakes up components when specific conditions are triggered. (2) Akita combines idle and congestion management in one mechanism, while HAsim only handles congestion. (3) Akita’s Availability Backpropagation is best suited to work with an event-driven simulation engine, while HAsim is better for a cycle-based FPGA environment.

\section{Conclusion}

The rapid proliferation of computer architecture simulators has created an urgent need to move beyond ad-hoc development toward systematic infrastructure that prioritizes both technical capability and usability. Akita addressed this challenge by demonstrating that a dedicated simulation engine, cleanly separated from architectural models, could deliver both performance and developer-friendliness through features like Smart Ticking, transparent parallel simulation, and comprehensive tracing support. Looking forward, we envision a future where simulation infrastructure becomes a shared community asset where architectural innovations are immediately composable across projects, where debugging and performance analysis are built-in rather than bolted-on, and where researchers spend their time exploring novel architectures rather than reimplementing common infrastructure. While Akita represents just one step toward this vision, we believe that elevating usability to equal importance with accuracy and performance will fundamentally transform how our community builds and shares simulation tools, ultimately accelerating the pace of computer architecture innovation.

\begin{acks}
We thank the National Science Foundation for its support under awards OAC-2246035 and OAC-2441804. Since the Akita simulation engine was originally extracted from MGPUSim and NaviSim, we thank the authors of the MGPUSim~\cite{sun2019mgpusim} and NaviSim~\cite{bao2022navisim} papers. We also thank the open-source contributors to the Akita and MGPUSim projects, including those who raised issues in the repositories. Any opinions, findings, and conclusions or recommendations expressed in this material are those of the authors and do not necessarily reflect the views of the National Science Foundation.
\end{acks}

\bibliographystyle{ACM-Reference-Format}
\bibliography{sample-base}

% \appendix









\end{document}